\documentclass[prb,aps,twocolumn,superscriptaddress,showpacs]{revtex4}

\usepackage{amsmath,amssymb,graphicx}
\usepackage{pxfonts}
\usepackage{graphicx}

\newcommand{\unit}[1]{~\mathrm{#1}} % for specifying units in math mode
\newcommand{\sub}[1]{$_\text{#1}$} % subscript in text mode

\begin{document}

\title{Pattern Formation in the Exciton Inner Ring}

\author{M.~Remeika}
\email{remeika@physics.ucsd.edu}
\author{A.T.~Hammack}
\affiliation{Department of Physics, University of California at San Diego, La Jolla, CA 92093-0319, USA}

\author{S.~Poltavtsev}
\affiliation{Department of Physics, University of California at San Diego, La Jolla, CA 92093-0319, USA}
\affiliation{Spin Optics Laboratory, St-Petersburg State University, St-Petersburg, Russia}

\author{L.V.~Butov}
\affiliation{Department of Physics, University of California at San Diego, La Jolla, CA 92093-0319, USA}

\author{J.~Wilkes}
\affiliation{Department of Physics and Astronomy, Cardiff University, Cardiff CF24 3AA, United Kingdom}

\author{A.L.~Ivanov}
\affiliation{Department of Physics and Astronomy, Cardiff University, Cardiff CF24 3AA, United Kingdom}

\author{K.L.~Campman}
\author{M.~Hanson}
\author{A.C.~Gossard}
\affiliation{Materials Department, University of California at Santa Barbara, Santa Barbara, CA 93106-5050, USA}

\begin{abstract}
We report on the spatially separated pump-probe study of indirect excitons in the inner ring in the exciton emission pattern. A pump laser beam generates the inner ring and a weaker probe laser beam is positioned in the inner ring. The probe beam is found to suppress the exciton emission intensity in the ring. We also report on the inner ring fragmentation and formation of multiple rings in the inner ring region. These features are found to originate from a weak spatial modulation of the excitation beam intensity in the inner ring region. The modulation of exciton emission intensity anti-correlates with the modulation of the laser excitation intensity. The three phenomena - inner ring fragmentation, formation of multiple rings in the inner ring region, and emission suppression by a weak probe laser beam - have a common feature: a reduction of exciton emission intensity in the region of enhanced laser excitation. This effect is explained in terms of exciton transport and thermalization.

\end{abstract}

\date{\today}

\maketitle

\section{Introduction}
\label{sec:introduction}

An exciton is a bound, typically hydrogen-like, state of an electron and a hole in a semiconductor. An indirect exciton is an exciton with an electron and a hole confined in spatially separated quantum wells layers \cite{Lozovik76}. Indirect excitons can be realized in a coupled quantum well structure (CQW) \cite{Fukuzawa90}. The spatial separation reduces the overlap of electron and hole wave functions thus allowing to achieve long exciton lifetimes, orders of magnitude longer than lifetimes of spatially direct excitons. In addition, indirect excitons are oriented dipoles with a built-in dipole moment $ed$, where $e$ is the electron charge and $d$ is the separation between the electron and hole layers (close to the distance between QW centers). The repulsive dipole-dipole interaction between indirect excitons allows them to screen in-plane disorder of the sample \cite{Ivanov02}. The combination of long lifetimes with a small amplitude of screened disorder allows indirect excitons to travel over large distances before recombination~\cite{Hagn95, Butov98, Larionov00, Butov02, Voros05, Ivanov06, Hammack06, Gartner06, Gartner07, Hammack07, High07, High08, Vogele09, Hammack09, Remeika09, Grosso09, High09, Kuznetsova10, Cohen11, Alloing11, Winbow11, Gorbunov11, Alloing12, Alloing12a, Kuznetsova12, Remeika12, Leonard12}. Furthermore, the built-in dipole moment allows to control indirect excitons by voltage $V_{\rm g}$ applied between electrodes above and below the QW layers: an electric field perpendicular to the QW plane $F_{\rm z}$ results in exciton energy shift by $edF_{\rm z} \propto V_{\rm g}$. The long lifetimes also allow indirect excitons to cool down to low temperatures below the temperature of quantum degeneracy \cite{Butov01}. These properties of indirect excitons - their ability to travel over large distances, their ability to cool to low temperatures, the possibility to control their transport by voltage and light, and the possibility to measure their transport by optical imaging - make indirect excitons a model system for studying transport of cold bosons in materials.

Transport of indirect excitons was studied in a variety of potential landscapes created by applied electric fields, including ramps \cite{Hagn95, Gartner06, Gartner07, Leonard12}, traps \cite{High09}, lattices \cite{Remeika09, Remeika12}, moving lattices—-conveyers \cite{Winbow11}, narrow channels \cite{Vogele09, Grosso09, Cohen11}, and circuit devices \cite{High07, High08, Grosso09, Kuznetsova10}, as well as in optically induced traps \cite{Hammack06, Hammack07, Gorbunov11, Alloing12a}. A set of exciton transport phenomena was observed, including the transistor effect for excitons \cite{High07, High08, Grosso09, Kuznetsova10}, localization-delocalization transition in random potentials \cite{Butov02, Ivanov06, Hammack06, Hammack07, Hammack09, Alloing11} and in static and moving lattices \cite{Remeika09, Winbow11, Remeika12}, and the inner ring in emission patterns \cite{Butov02, Ivanov06, Stern08, Hammack09, Remeika09, Alloing12, Remeika12, Kuznetsova12}. The studies of the latter form the subject of this work.

Long-lived excitons generated by focused excitation can form an emission ring around the excitation spot, referred to as the inner ring. The inner ring occurs due to the heating of the exciton gas by laser excitation. The heating results in a reduction in the occupation of the low-energy optically active exciton states \cite{Feldmann87, Hanamura88, Andreani91} and, in turn, the exciton emission intensity. When excitons travel away from the excitation spot they thermalize to the lattice temperature so that the emission intensity increases outside of the excitation spot forming the inner ring \cite{Butov02, Ivanov06, Hammack09, Remeika09, Alloing12, Remeika12, Kuznetsova12}.

Besides the inner ring, the external ring can be observed around the excitation spot \cite{Butov02, Butov04, Rapaport04, Chen05, Haque06, Yang06, Yang07, Yang10, High12, Alloing12b} and localized bright spot (LBS) rings can be observed around localized carrier sources \cite{Butov02, Butov04, Lai04, Yang06, Yang07, Yang10, High12}. The external and LBS rings form on the boundaries between electron-rich and hole-rich regions; the former is created by current through the structure and the latter is created by optical excitation \cite{Butov04, Rapaport04, Chen05, Haque06, Yang10}.

At low temperatures, intensity modulation \cite{Butov02, Butov04, Yang06, Yang07, Yang10, High12, Alloing12b} and spontaneous coherence of excitons \cite{Yang06, High12, Alloing12b} are observed in the external ring; intensity modulation \cite{Butov04} and spontaneous coherence of excitons \cite{High12} are also observed in the LBS rings. The exciton state with the spatial order of higher-intensity exciton beads on a macroscopic length scale (up to $\sim$mm) is referred to as the macroscopically ordered exciton state (MOES) \cite{Butov02}. The exciton coherence length in the MOES reaches $\sim 10$ $\mu$m and is much larger than the coherence length, which can be achieved in a classical gas, showing that the MOES is a condensate in momentum space \cite{Yang06, High12}. The development of understanding of the MOES is in progress. Theoretical models describing various instabilities in the external ring will be reviewed elsewhere.

In contrast to the external and LBS rings, the intensity profile along the inner ring was smooth \cite{Ivanov06} and no intensity modulation was reported in the inner ring until this work. In this article, we present observations of radial and azimuthal modulation of the emission intensity of indirect excitons in the inner ring. We also present a spatially separated pump-probe study of indirect excitons in the inner ring: A probe beam positioned in the inner ring generated by a pump beam is found to suppress the exciton emission intensity. The inner ring fragmentation, formation of multiple rings in the inner ring region, and emission suppression by a probe laser beam are all characterized by a reduction of exciton emission intensity in the region of enhanced laser excitation. We discuss the origin of these phenomena and present a model which explains a reduction of exciton emission intensity in the region of enhanced laser excitation.

In Section II, we describe the spatially separated pump-probe study of indirect excitons in the inner ring. In Section III, we present the radial and azimuthal modulation of the emission intensity of indirect excitons in the inner ring. A summary of the work is given in Section IV. In Appendix, we show that the reported fragmented inner ring is qualitatively different from the earlier observed fragmented external ring.

\section{Spatially-resolved pump-probe study}
\label{sec:pump-probe}

\subsection{Experiments}

\begin{figure}[b]
\centering
\includegraphics[width=7.5cm]{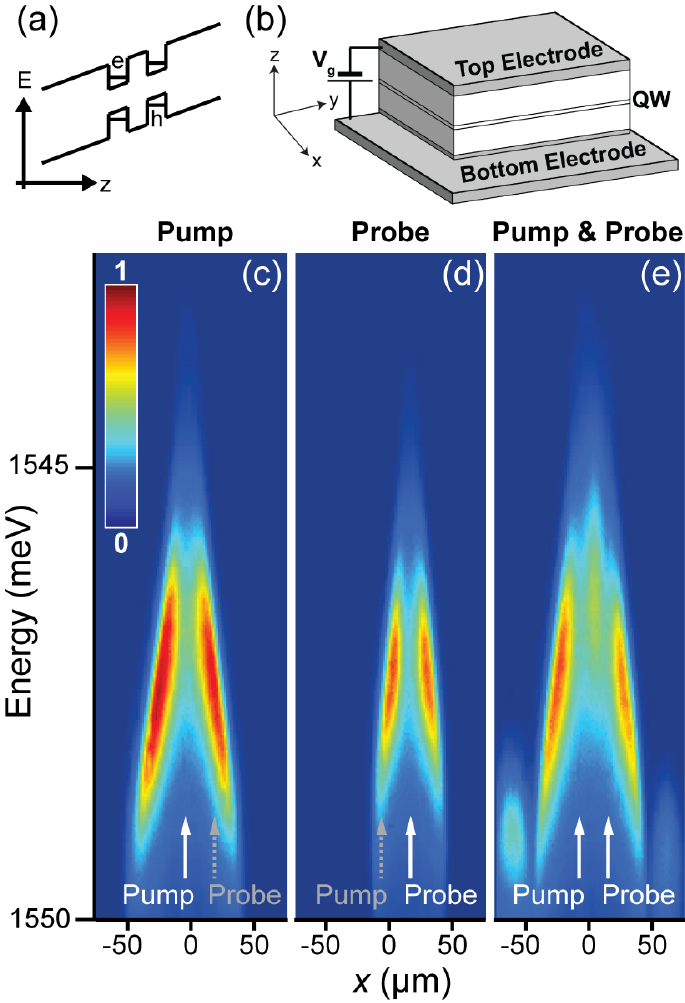}
\caption{Exciton emission images. (a) CQW band diagram. (b) Sample structure. GaAs quantum wells are positioned within the insulating Al\sub{0.33}Ga\sub{0.67}As layer (white) surrounded by conducting $n^+-$GaAs layers (gray). Voltage $V_{\rm g}$ is applied between top and bottom electrodes to create an electric field perpendicular to the QW plane. (c-e) $x-$energy images of the emission of indirect excitons with pump (c), probe (d), and both (e) beams present. The pump and probe laser power $P_{\rm pump}=700\unit{\mu W}$, $P_{\rm probe}=100\unit{\mu W}$ and energy $E_{\rm pump}=E_{\rm probe}=1960\unit{meV}$. The beam positions are indicated by arrows.}
\label{fig:1}
\end{figure}

\begin{figure}
\centering
\includegraphics[width=7.5cm]{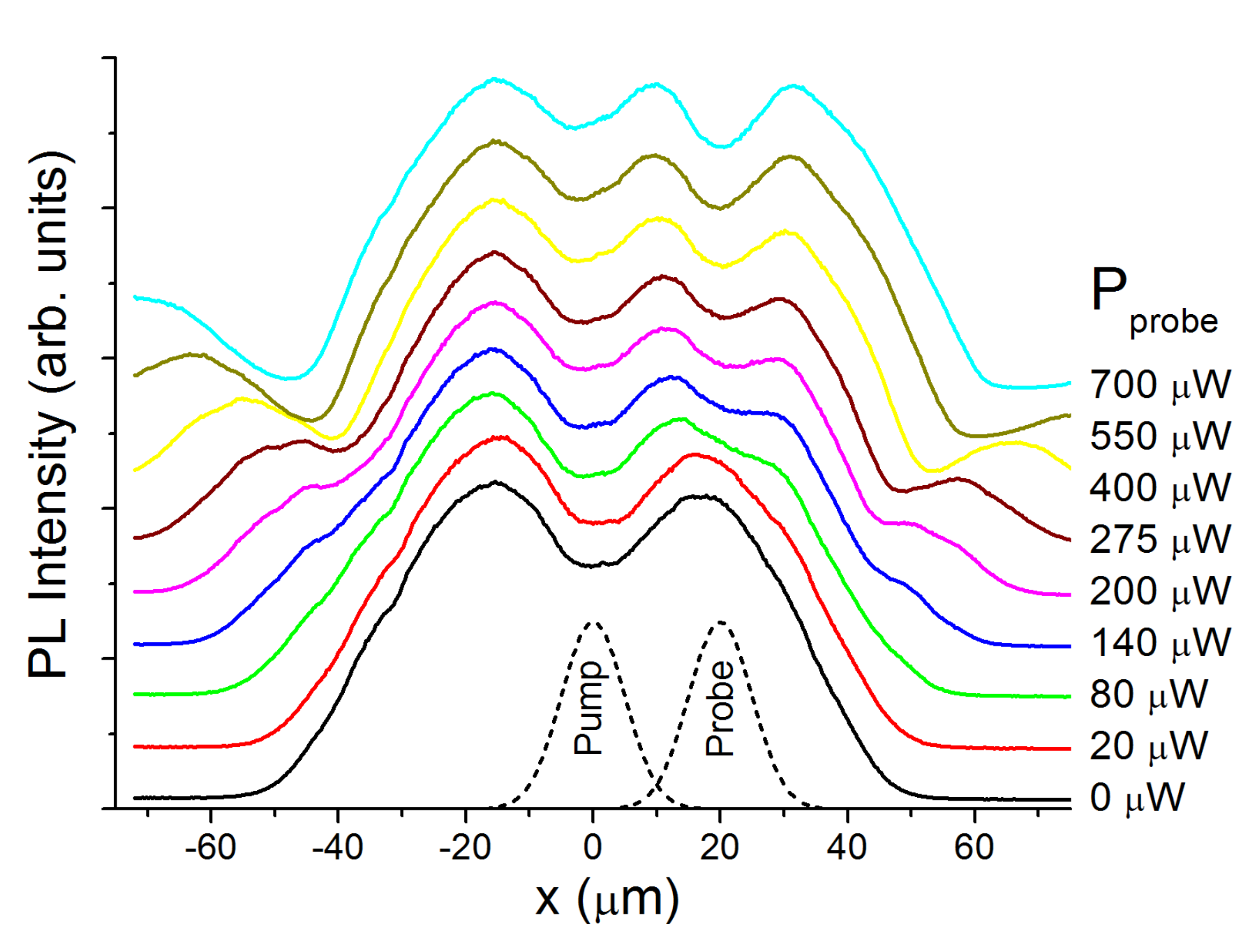}
\caption{Exciton emission profiles in the spatially-resolved pump-probe experiment. Measured emission intensity profiles of indirect excitons for increasing $P_{\rm probe}$. Plots are offset for clarity. The pump and probe excitation profiles are shown by dashed lines. The pump beam generates the inner ring and the probe beam is positioned in the inner ring. The external ring can also be seen beyond the inner ring, the external ring radius (that is the distance to the excitation line for the studied case of line excitation) quickly increases with the excitation power. $E_{\rm pump}=E_{\rm probe}=1960\unit{meV}$. Pump power $P_{\rm pump}=700\unit{\mu W}$ and temperature $T = 1.6\unit{K}.$}
\label{fig:2}
\end{figure}

A common feature of pump-probe experiments is to generate an excitation using a ``pump'' beam and to probe the system using a (weaker) ``probe'' beam. Typically, the pump and probe excitations are separated in time \cite{Schmitt-Rink89}. Here, we present the measurements where the pump and probe laser beams excite simultaneously and continuously but are separated in space: A pump laser beam generates the inner ring and a weaker probe laser beam is positioned in this inner ring. We study the effect of a probe beam on an exciton cloud generated by a pump beam. Furthermore, as described in Section~\ref{sec:patterns}, these spatially resolved pump-probe experiments clarify the origin of the inner ring fragmentation.

The CQW structure is grown by molecular beam epitaxy (MBE). The structure consists of two GaAs quantum wells separated by $4\unit{nm}$ Al\sub{0.33}Ga\sub{0.67}As barrier and surrounded by $200\unit{nm}$ Al\sub{0.33}Ga\sub{0.67}As barrier layers. The top and bottom electrodes are Si-doped $n^+-$GaAs layers with $n_{Si} = 5\times 10^{17}\unit{cm}^{-3}$. The voltage $V_{\rm g}$ applied between the conducting $n^+-$GaAs layers drops in the insulating layer between them (Fig.~\ref{fig:1}a,b), $V_{\rm g}=1.4\unit{V}$ for the data presented in this Section. The details on this sample can be found in Ref. \cite{Butov99}.

A $632\unit{nm}$ HeNe laser is used for a pump beam and a tunable-frequency Ti:Sapphire laser or $632\unit{nm}$ HeNe laser is used for a probe beam. The excitation geometry is linear, with both the pump and probe beams focused into lines greater than $200\unit{\mu m}$ long and with full width at half-maximum ${\rm FWHM}\approx 10\unit{\mu m}$. In this geometry, the inner ring is observed as a pair of lines on each side of the laser excitation line. The pump and probe lasers are operated in cw mode. Although the line geometry is different from the ring geometry, for clarity, we keep calling the regions of enhanced emission intensity around the excitation spot the inner ring. Emission images are captured by a CCD camera with a bandpass filter selecting photon wavelengths $\lambda = 800 \pm 5\unit{nm}$ covering the spectral range of the indirect exciton emission. Spectra are measured using a spectrometer with resolution $0.18\unit{meV}$. The measurements are done in a helium cryostat at $1.6\unit{K}$.

For a weak laser excitation, the density of optically generated excitons is low, they are localized by the in-plane disorder potential in the structure and their emission profile essentially follows the laser excitation profile \cite{Butov02, Ivanov06, Hammack09, Remeika09, Alloing12, Remeika12, Kuznetsova12}. For a strong laser excitation, the density of generated excitons is high so that they can effectively screen the disorder potential, delocalize, and travel and cool to the lattice temperature outside the laser excitation spot thus forming the inner ring \cite{Butov02, Ivanov06, Hammack09, Remeika09, Alloing12, Remeika12, Kuznetsova12}.

\begin{figure}
\centering
\includegraphics[width=7.5cm]{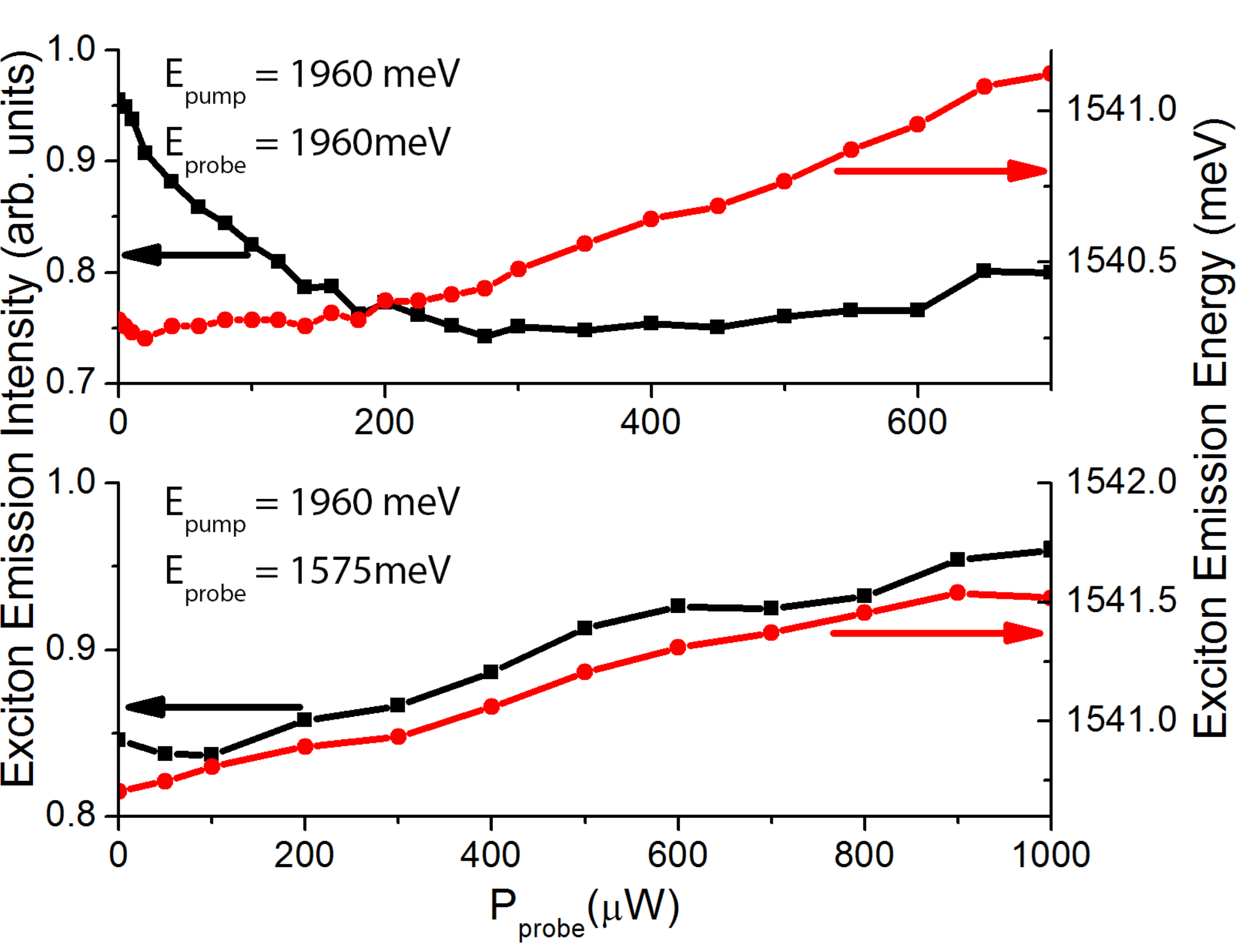}
\caption{Effect of probe beam on indirect exciton emission. (a) High-energy probe beam ($E_{\rm probe}=1960\unit{meV}$) positioned in the inner ring generated by the pump beam causes the emission suppression of indirect excitons (black). The energy of indirect excitons (red) monotonically increases with the probe power $P_{\rm probe}$. (b) Probe beam tuned to the direct exciton energy ($E_{\rm probe}=1575\unit{meV}$) does not suppress the indirect exciton emission. Both the emission intensity (black) and energy (red) of indirect excitons monotonically increase with $P_{\rm probe}$. $E_{\rm pump}=1960\unit{meV}$ and $P_{\rm pump}=700\unit{\mu W}$ both for (a) and (b).}
\label{fig:3}
\end{figure}

Figures~\ref{fig:1}c-e show the examples of $x-$energy emission images of indirect excitons. Figure~\ref{fig:1}c shows the inner ring generated by the pump laser. In all experiments presented in this Section, the energy of the pump laser $E_{\rm pump}$ is considerably ($\sim 400$ meV) higher than the exciton energy. Such high-energy laser excitation strongly heats excitons in the excitation spot and can create a pronounced inner ring pattern \cite{Kuznetsova12}. Figure~\ref{fig:1}d shows the emission pattern generated by a weaker probe laser. A similar ring pattern is seen as in Fig.~\ref{fig:1}c with a smaller inner ring radius due to a lower density of generated excitons. These single-beam patterns are consistent with the patterns measured in earlier studies \cite{Butov02, Ivanov06, Hammack09, Remeika09, Alloing12, Remeika12, Kuznetsova12}. Figure~\ref{fig:1}e shows the emission pattern when both pump and probe beams are present. The probe beam is positioned in the inner ring generated by the pump beam. Figure~\ref{fig:1}e shows that the addition of the probe beam in the inner ring reduces the emission intensity in the probe beam location. This counterintuitive behavior is detailed in Fig.~\ref{fig:2} and \ref{fig:3} and described below.

Figure~\ref{fig:2} shows a series of emission profiles of indirect excitons for fixed pump laser power $P_{\rm pump}$ and varying probe laser power $P_{\rm probe}$. For $P_{\rm probe}=0$, the pump beam generates the inner ring around it (the lowest profile in Fig.~\ref{fig:2}). An increase of the probe laser power results in the suppression of the exciton emission in the probe beam location (Fig.~\ref{fig:2}). The emission suppression can be also seen in Fig.~\ref{fig:3}a, which presents the emission intensity of indirect excitons in the probe beam location. Note that a weak probe beam with $P_{\rm probe} < 100\unit{\mu W}$ causes a significant suppression of the emission intensity in the inner ring generated by an order of magnitude stronger $P_{\rm pump}=700\unit{\mu W}$ pump beam.

Figure~\ref{fig:3} also shows that the exciton emission energy monotonically increases with the probe power $P_{\rm probe}$. The repulsive interaction between indirect excitons in the CQW results in the exciton energy enhancement with density \cite{Butov99, Butov02, Ivanov06, Hammack06, Yang07, Hammack09, Remeika09, High09, Yushioka99, Zhu95, Lozovik96, Laikhtman01, Schindler08, Ivanov10}. The observed energy enhancement (Fig.~\ref{fig:3}) indicates that the exciton density increases with increasing $P_{\rm probe}$. Figure~\ref{fig:3}a shows that the high-energy probe beam suppresses the emission intensity of indirect excitons in the probe beam location, while increasing the density of indirect excitons.

Figure~\ref{fig:3}b presents the data for the probe beam energy $E_{\rm probe}$ tuned to the direct exciton resonance: $E_{\rm probe}=1575\unit{meV}$. As shown in previous studies \cite{Kuznetsova12}, such resonant excitation minimizes the excitation-induced heating of indirect excitons. Figure~\ref{fig:3}b shows that the resonant probe beam increases both emission intensity and density of indirect excitons.

These data suggest that the high-energy probe beam results in substantial local heating of the exciton gas that lowers the occupation of the low-energy optically active exciton states and, in turn, the exciton emission intensity even in spite of the increase of the total local exciton density. This interpretation is supported by the simulations described in the next Section.

\subsection{Simulations}
\label{sec:theory}

\begin{figure}
\centering
\includegraphics[width=7.5cm]{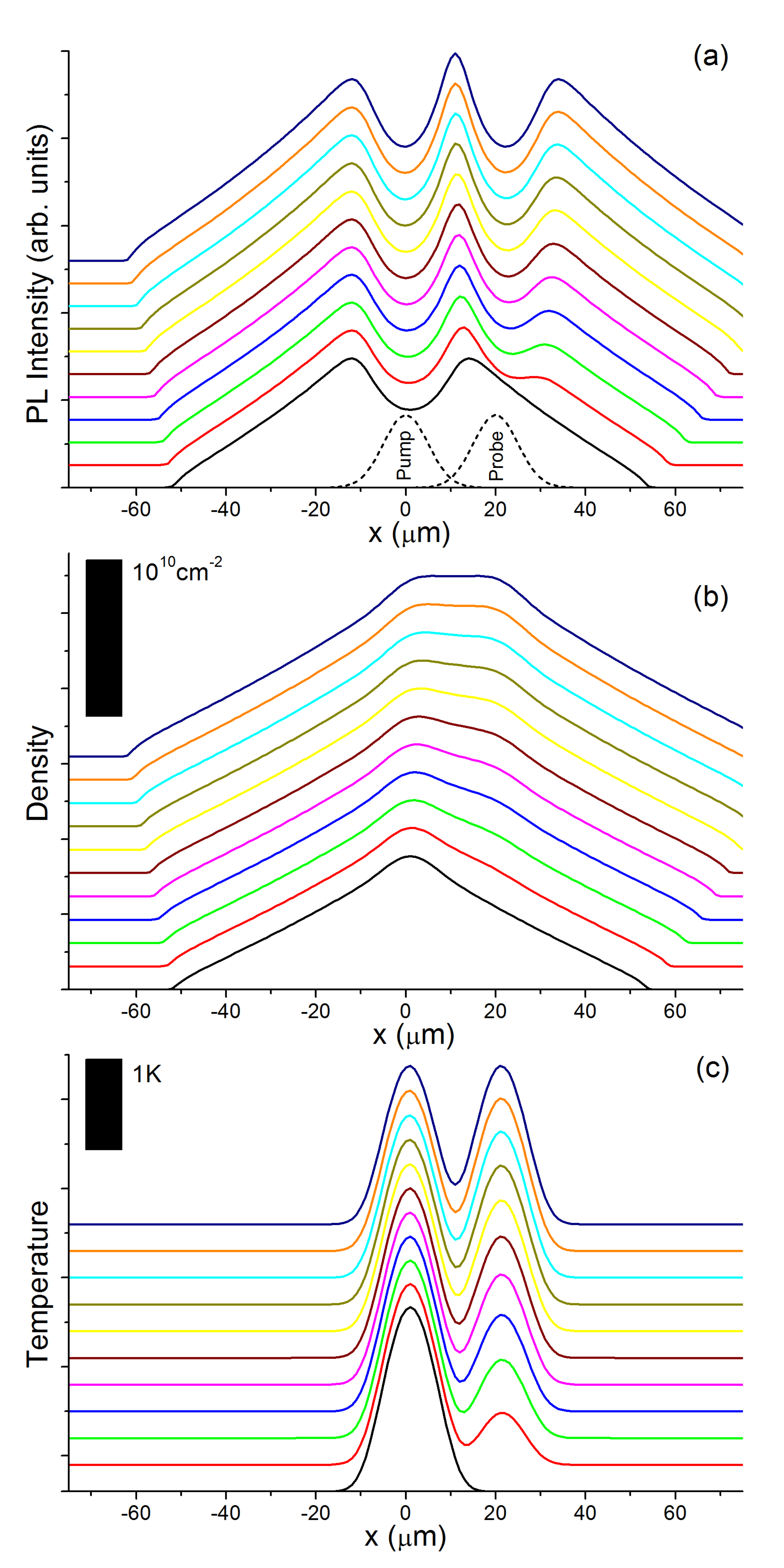}
\caption{Exciton emission profiles in the spatially-resolved pump-probe study. (a-c) Calculated profiles of emission intensity (a), density (b), and temperature (c) of indirect excitons for increasing $\Lambda_{\rm probe}$. The pump generation rate $\Lambda_{\rm pump}=8 \times 10^8\unit{cm^{-2} ns^{-1}}$. The probe generation rate increases from $\Lambda_{\rm probe}=0$ (bottom) to $\Lambda_{\rm probe}=8 \times 10^8\unit{cm^{-2} ns^{-1}}$ (top). Plots are offset for clarity. The pump and probe excitation profiles are shown by dashed lines. The pump beam generates the inner ring and the probe beam is positioned in the inner ring. The temperature corresponding to flat parts of the curves in (c) is 1.6 K.}
\label{fig:4}
\end{figure}

A model based on the transport, thermalization and optical decay of indirect excitons \cite{Ivanov06, Hammack09} is used to simulate the exciton spatial photoluminescence (PL) patterns. Full details of the model can be found in Ref.~\cite{Hammack09}. This Section gives a brief schematic of the model and describes details of the pump-probe simulations relevant to the experiments presented above. The steady-state density distribution of dipole-oriented indirect excitons, $n_{\rm x}$, satisfies the following non-linear transport equation,
\begin{equation}
\nabla [D \nabla n_{\rm x} + \mu n_{\rm x} \nabla (u_0 n_{\rm x})] - \Gamma_{\rm opt} n_{\rm x} + \Lambda = 0.
\label{transport}
\end{equation}
For the line excitation geometry of the pump-probe experiments where the system is homogeneous in the $y$-direction, $\nabla = \partial/\partial x$ is used. The two terms in the square brackets of Eq.~(\ref{transport}) are due to the diffusion and drift currents of excitons, respectively. The density and temperature dependence of the diffusion coefficient $D = D_0 {\rm exp}[-U_0/(u_0 n_{\rm x} + k_B T)]$ describes the screening of the QW disorder potential by the repulsively interacting indirect excitons \cite{Ivanov02}. $u_0 n_{\rm x}$ is the dipole-dipole interaction potential and $U_0$ is the amplitude of the disorder potential. The exciton mobility is given by the generalized Einstein relation, $\mu = D(e^{T_0/T} - 1)/(k_B T_0)$. Here, $T_0 = (\pi \hbar^2 n_{\rm x})/(2 M_{\rm x} k_B)$ is the temperature of quantum degeneracy ($M_{\rm x}=0.22m_0$ is the exciton mass). The optical decay rate $\Gamma_{\rm opt}(T_0,T)$ includes the fact that only excitons in the low energy, optically active states may emit light \cite{Feldmann87, Hanamura88, Andreani91}. The exciton generation rate $\Lambda = \Lambda_{\rm pump} + \Lambda_{\rm probe}$ has contributions from the pump and probe beams, $\Lambda_{\rm pump}$ and $\Lambda_{\rm probe}$, respectively. These have Gaussian profiles with position and FWHM chosen to match the experiments.

The profile of the indirect exciton effective temperature $T$, which stands in the expressions for the diffusion coefficient, mobility, and optical decay rate, is found by solving a heat balance equation,
\begin{eqnarray}
S_{\rm phonon}(T_0,T) &= S_{\rm laser}(T_0,T,\Lambda_{\rm pump},E^{({\rm ex})}_{\rm pump}) + \\*
											& +S_{\rm laser}(T_0,T,\Lambda_{\rm probe},E^{({\rm ex})}_{\rm probe}).\nonumber
\label{thermalization}
\end{eqnarray}
Here, $S_{\rm phonon}$ is the rate of cooling of indirect excitons towards the lattice temperature by a bath of bulk longitudinal acoustic phonons. In the case of non-resonant excitation, heating of the exciton gas occurs due to the injection of high-energy excitons, which then thermalize within a time scale much shorter than the cooling time associated with acoustic phonons \cite{Ivanov99}. The laser induced heating rate $S_{\rm laser}(T_0,T,\Lambda_{\rm pump(probe)},E^{({\rm ex})}_{\rm pump(probe)})$ is characterized by the injection rate $\Lambda_{\rm pump(probe)}$ and the excess energy $E^{({\rm ex})}_{\rm pump(probe)}$ acquired by photoexcited excitons due to the pump(probe) beams. The excess energy relates to the laser excitation energy $E_{\rm pump(probe)}$ by $E^{({\rm ex})}_{\rm pump(probe)} = E_{\rm pump(probe)} - E_{IX}$ where $E_{IX}$ is the indirect exciton energy. Expressions for $S_{\rm phonon}$, $S_{\rm laser}$, $\Gamma_{\rm opt}$ and all other parameters of the model are given in Ref.\,\cite{Hammack09}.

Solving numerically Eqs.~(\ref{transport}-\ref{thermalization}), reveals the spatial profiles of the density $n_{\rm x}$, temperature $T$ and optical decay rate $\Gamma_{\rm opt}$ of indirect excitons. The emission intensity $I_{PL} = \Gamma_{\rm opt} n_{\rm x}$ is used to make comparisons with the experimental data.

The results of the simulations are presented in Fig.~\ref{fig:4}. They show that illuminating the inner ring with a weaker probe beam produces local heating of the exciton gas (Fig.~\ref{fig:4}c) and, as a result, a darkening of the exciton emission in that area (Fig.~\ref{fig:4}a) even in spite of the increase of the local exciton density (Fig.~\ref{fig:4}b), in agreement with the experimental data (Fig.~\ref{fig:2}).

\section{Azimuthal and Radial Fragmentation of the Inner Ring}
\label{sec:patterns}

\begin{figure}
\centering
\includegraphics[width=5.5cm]{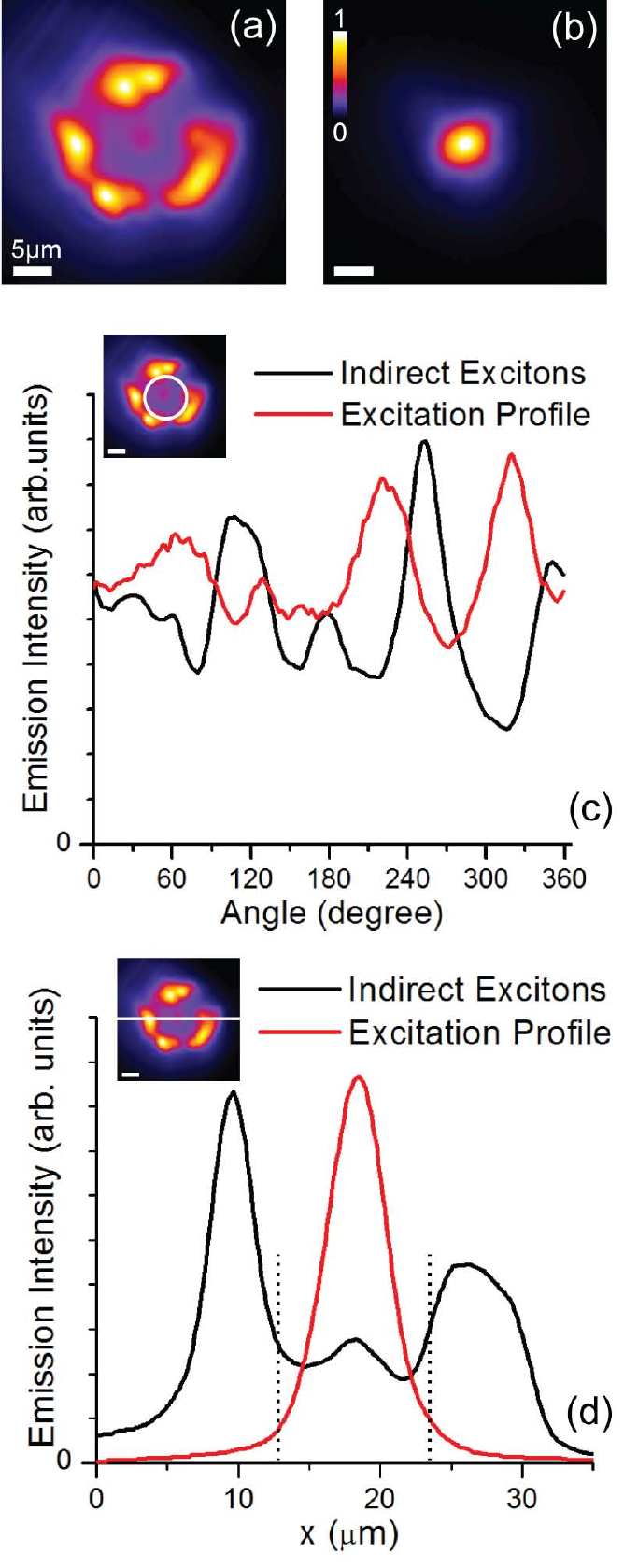}
\caption{Azimuthal fragmentation of the inner ring. (a) Spatial distribution of emission intensity of indirect excitons. (b) Laser excitation shape. (c) Normalized azimuthal profiles of emission intensity of indirect excitons (black) and laser excitation (red) in the inner ring region. The profiles are measured along the circular path around the laser excitation center, see inset. The modulation of exciton emission intensity anti-correlates with the modulation of the laser excitation intensity. (d) Radial intensity profiles of emission intensity of indirect excitons (black) and laser excitation (red). The profiles are measured along the straight line through the laser excitation center, see inset. The dotted lines in (d) show the location of the circular path in (c). A bump in the center seen in (a) and (d) is due to GaAs bulk emission in the spectral range of indirect excitons. The laser excitation energy $E=1960\unit{meV}$ and power $P=10\unit{\mu W}$.}
\label{fig:5}
\end{figure}

\begin{figure}
\centering
\includegraphics[width=5.5cm]{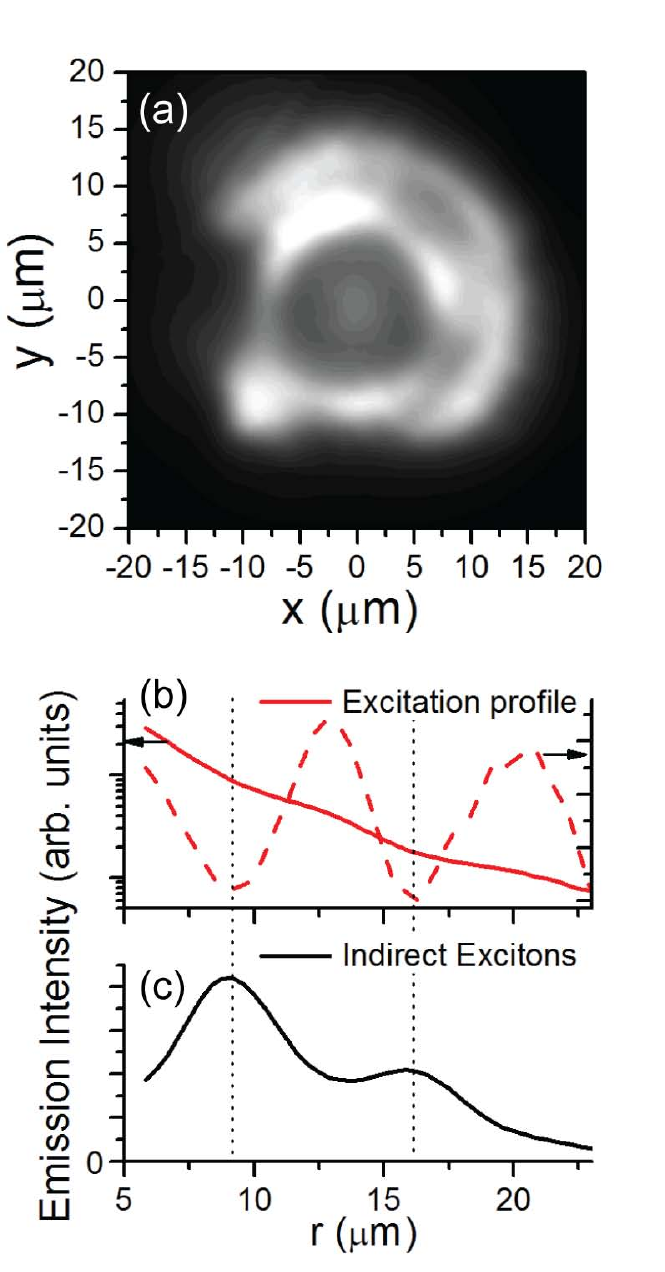}
\caption{Radial fragmentation of the inner ring. (a) Spatial distribution of emission intensity of indirect excitons. Left and bottom edges are cut off at the electrode edge. (b) Radial profile of laser excitation (solid line). Dashed line shows negative second derivative of logarithm of intensity to magnify the intensity modulation on a slowly varied background. (c) Radial profiles of emission intensity of indirect excitons. Dotted lines show the anti-correlation between the laser excitation (b) and exciton emission intensity (c). The laser excitation energy $E=1960\unit{meV}$ and power $P=8\unit{\mu W}$.}
\label{fig:6}
\end{figure}

In this Section we present azimuthal and radial fragmentation of the inner ring. The CQW structure used in these studies is similar to the structure used in the pump-probe experiment described in Section~\ref{sec:pump-probe}. Two $8\unit{nm}$ GaAs quantum wells separated by $4\unit{nm}$ Al\sub{0.33}Ga\sub{0.67}As barrier, are enclosed between $100\unit{nm}$ (bottom) and $900\unit{nm}$ (top) Al\sub{0.33}Ga\sub{0.67}As barrier layers. The bottom electrode is Si-doped $n^+-$GaAs layer with $n_{Si} = 10^{18}\unit{cm}^{-3}$ and the top electrode is a semitransparent sputter-deposited $80\unit{nm}$ ITO layer. The voltage $V_{\rm g}$ applied between the top and bottom electrodes drops in the insulating layer between them (Fig.~\ref{fig:1}a,b). For samples with the same layer structure, designed patterns of top electrodes create the required in-plane potential landscapes for excitons, including traps \cite{High09}, lattices \cite{Remeika09, Winbow11, Remeika12}, ramps \cite{Leonard12}, and circuit devices \cite{High07, High08, Grosso09, Kuznetsova10}. In this work, we use a top electrode unstructured on a large area. Such electrode creates a laterally homogeneous $F_{\rm z}$ as in the experiments in Section~\ref{sec:pump-probe}. Excitons are generated by a 700 nm Ti:Sapphire laser or $632\unit{nm}$ HeNe laser focused to the excitation spot $\sim 5 \mu$m in diameter. Imaging and spectroscopy are performed the same way as in the experiment described in Section~\ref{sec:pump-probe}.

Figure~\ref{fig:5}a shows a fragmented exciton inner ring. The azimuthal variations of exciton emission intensity anti-correlate with the azimuthal variations of the tails of laser excitation (Fig.~\ref{fig:5}c). This suggests that the origin of the inner ring fragmentation is the exciton emission suppression by the high-energy laser excitation in the inner ring region. A mechanism of exciton emission suppression in the inner ring by high-energy laser excitation is described in Section~\ref{sec:pump-probe}.
Note that the intensity of the laser tails that cause the fragmentation is too low to be visible in the same image as the intensity peak at the center of the spot (Fig. \ref{fig:5}b).

Another pattern in the exciton inner ring is shown in Fig.~\ref{fig:6}a. In this case, the laser spot is shaped to be surrounded by weak Airy rings. Airy rings originate from the beam diffraction and their radius and intensity can be controlled by optics \cite{HectOptics}. In this experiment, an iris aperture is introduced into the laser beam to create weak Airy rings of laser excitation in the inner ring region. Airy rings of laser excitation can result in the formation of multiple rings in exciton emission. Two rings can be clearly seen in the exciton emission pattern in the inner ring region (Fig.~\ref{fig:6}). This effect is caused by a weak radial modulation of the laser intensity (Fig.~\ref{fig:6}b). Figure~\ref{fig:6}b,c shows that bright Airy rings of laser excitation anti-correlate with bright rings of exciton emission. Both in the case of the azimuthal (Fig.~\ref{fig:5}) and radial (Fig.~\ref{fig:6}) fragmentation of the inner ring, a weak enhancement of the laser excitation intensity in the inner ring region, orders of magnitude weaker than the excitation used to generate the exciton cloud, produces a substantial suppression of exciton emission in the inner ring.

\begin{figure}
\centering
\includegraphics[width=7.5cm]{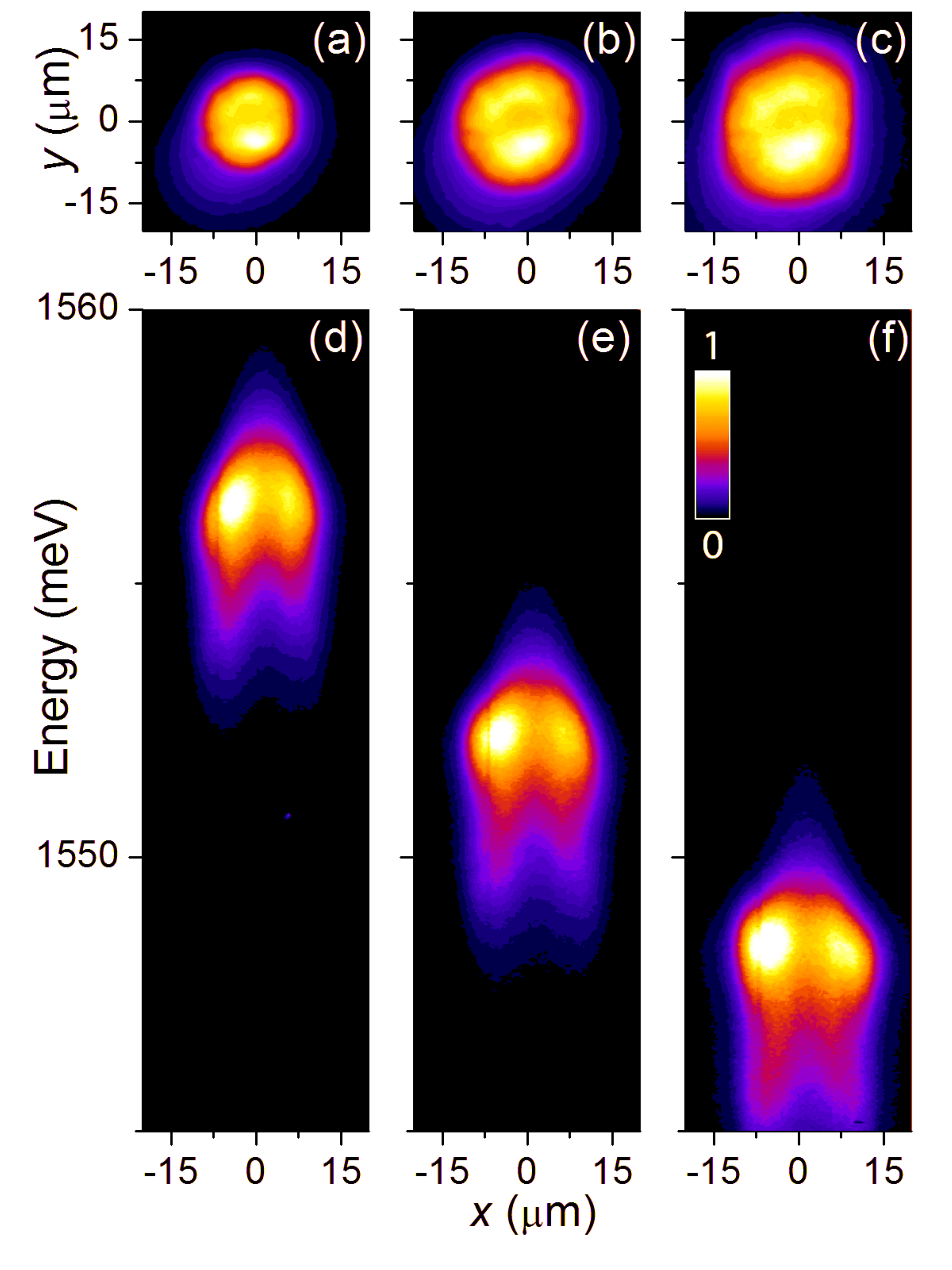}
\caption{Inner ring for a lower-energy Ti:Sapphire laser excitation. (a-c) Spatial images of the indirect exciton emission for the electrode voltage $V_{\rm g} = -2\unit{V}$ (a), $-2.5\unit{V}$ (b), and $-3\unit{V}$ (c). (d-f) $x-$energy images of the indirect exciton emission for the electrode voltage $V_{\rm g} = -2\unit{V}$ (d), $-2.5\unit{V}$ (e), and $-3\unit{V}$ (f). The laser excitation power $P=16\unit{\mu W}$ (a-c) and $10\unit{\mu W}$ (d-f), and energy $E=1771\unit{meV}$.}
\label{fig:7}
\end{figure}

No substantial fragmentation of the inner ring is observed in the case of a lower-energy Ti:Sapphire excitation (Fig.~\ref{fig:7}). This excitation does not cause as strong heating of the exciton gas as the high-energy excitation \cite{Kuznetsova12}. The contrast of the inner ring itself, i.e. the reduction of the exciton emission intensity in the laser excitation center, is also substantially weaker for such excitation for the same reason \cite{Kuznetsova12}. Furthermore, no substantial fragmentation of the inner ring is observed for the high-energy laser excitation when the tails of the laser excitation spot in the inner ring region are smooth, see Fig. 1d in Ref.~\cite{Ivanov06}. These data confirm that the observed azimuthal and radial fragmentation of the exciton emission intensity in the inner ring are caused by the heating of a cold exciton gas in the inner ring by a spatially modulated laser excitation in the inner ring region.

The fragmentation of the inner ring is clearly different from the fragmentation of the external and LBS rings studied earlier. While the fragmentation of the inner ring is determined by the shape of the laser excitation tails in the inner ring region, the fragmentation of the external and LBS rings forms spontaneously as described in \cite{Butov02, Butov04, Yang06, Yang07, High12, Alloing12b}, see Section~I. Details on the comparison of the fragmented inner ring and fragmented external ring are presented in Appendix.

\section{Summary}

We studied the effect of laser excitation on the inner ring in the exciton emission pattern. We performed the spatially separated pump-probe study of indirect excitons in the inner ring. A pump laser beam generates the inner ring and a weaker probe laser beam is positioned in the inner ring. The probe beam is found to suppress the exciton emission intensity in the ring. We also report on the inner ring fragmentation and formation of multiple rings in the inner ring region. These features are found to originate from a weak spatial modulation of the excitation beam intensity in the inner ring region. The modulation of exciton emission intensity anti-correlates with the modulation of the laser excitation intensity. The three phenomena - inner ring fragmentation, formation of multiple rings in the inner ring region, and emission suppression by a weak probe laser beam - have a common feature: a reduction of exciton emission intensity in the region of enhanced laser excitation. Another shared feature of all these patterns is that a weak enhancement of the laser excitation intensity in the inner ring region, orders of magnitude weaker than the excitation used to generate the exciton cloud, produces a substantial suppression of exciton emission in the inner ring. The observed fragmentation of the inner ring is clearly different from the fragmentation of the external and LBS rings studied earlier. A reduction of exciton emission intensity in the region of enhanced laser excitation is explained in terms of exciton transport and thermalization: The experimental data and theoretical simulations show that illuminating the inner ring with a high-energy laser excitation produces local heating of the exciton gas and, as a result, suppresses the exciton emission in that area in spite of the increase of the local exciton density.

\section{Acknowledgments}

We thank Marc Baldo, Alexey Kavokin, Yuliya Kuznetsova, Jason Leonard, Leonid Levitov, and Ben Simons for discussions. Support of this work by NSF and EPSRC is gratefully acknowledged.

\section{Appendix: Inner Ring vs External Ring}

\begin{figure}
\centering
\includegraphics[width=8.5cm]{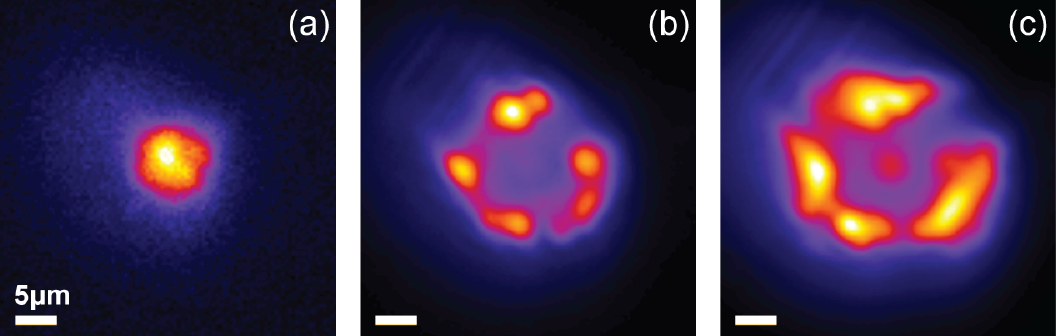}
\caption{The inner ring dependence on excitation power. (a-c) The inner ring with increasing excitation power: $P=4\unit{\mu W}$ (a), $9\unit{\mu W}$ (b), and $12\unit{\mu W}$ (c). The emission peak seen in the center of the ring in (b,c) is due to bulk GaAs emission in the spectral range of indirect excitons. $V_{\rm g}=3$~V, $E=1960$~meV.}
\label{fig:8}
\end{figure}

\begin{figure}
\centering
\includegraphics[width=8.5cm]{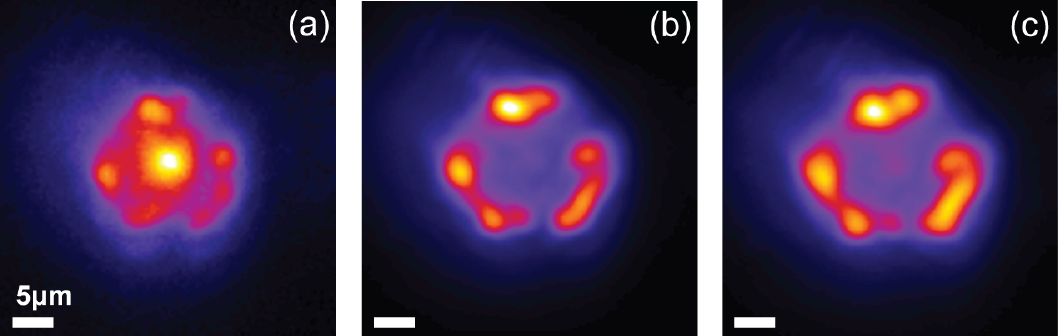}
\caption{The inner ring dependence on applied voltage. (a-c) The inner ring with increasing voltage: $V_{\rm g}=1.5\unit{V}$ (a), $2.5\unit{V}$ (b), and $3.5\unit{V}$ (c). The emission peak seen in the center of the ring is due to bulk GaAs emission in the spectral range of indirect excitons. $P=10\unit{\mu W}$, $E=1960\unit{meV}$. }
\label{fig:9}
\end{figure}

\begin{figure}
\centering
\includegraphics[width=8.5cm]{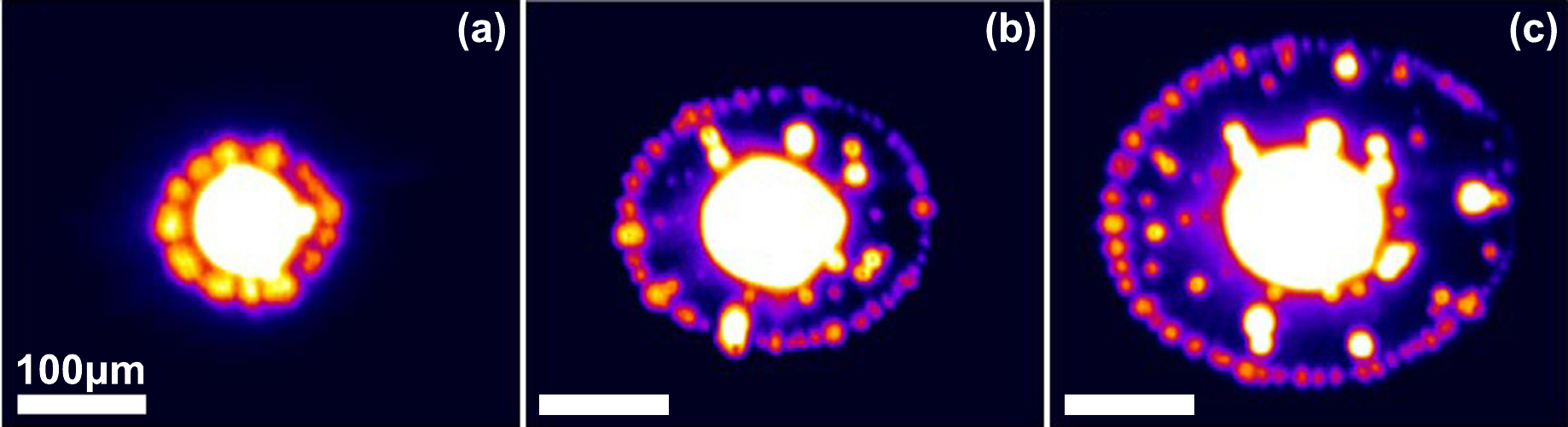}
\caption{The external ring dependence on excitation power. (a-c) The external ring with increasing excitation power: $P=310\unit{\mu W}$ (a), $560\unit{\mu W}$ (b), and $930\unit{\mu W}$ (c). $V_{\rm g}=1.24\unit{V}$, $E=1960\unit{meV}$.}
\label{fig:10}
\end{figure}

\begin{figure}
\centering
\includegraphics[width=8.5cm]{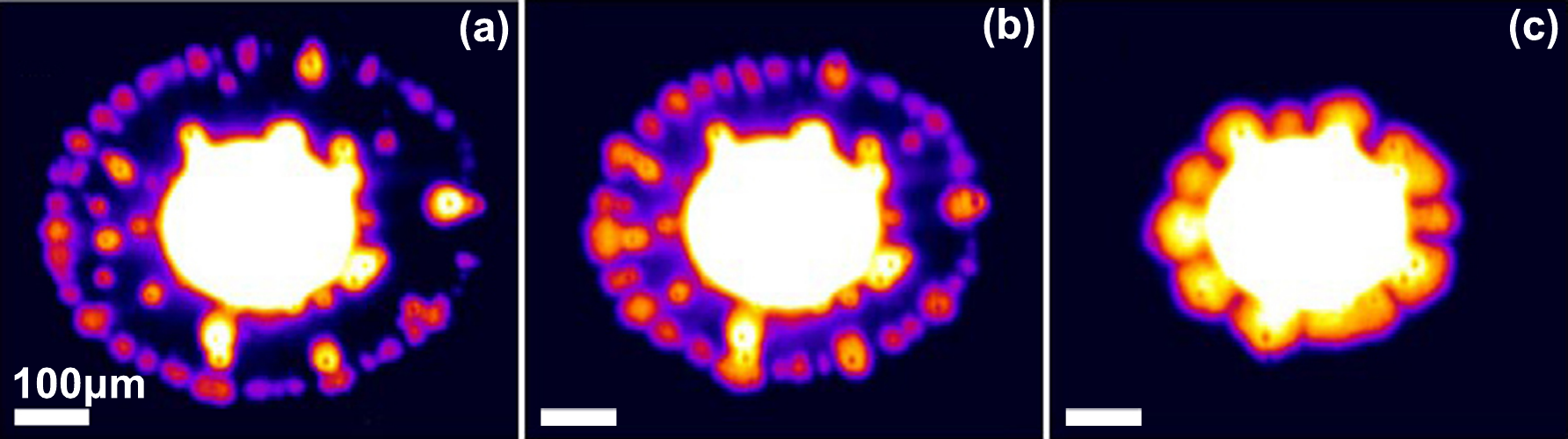}
\caption{The external ring dependence on applied voltage. (a-c) The external ring with increasing voltage: $V_{\rm g}=1.262\unit{V}$ (a), $1.274\unit{V}$ (b), and $1.3\unit{V}$ (c). $P=1400\unit{\mu W}$, $E=1960\unit{meV}$. }
\label{fig:11}
\end{figure}

Both the inner and external ring form around the laser excitation spot. However, the origin of the inner ring and the external ring is clearly different (see Section I). The origin of the inner ring fragmentation and the external ring fragmentation is also clearly different (see Sections I-III). It is important to distinguish the inner and external ring in the experiments. In this Appendix, we present the differences between the inner ring and external ring. In particular, the dependence of the ring on the laser excitation power and energy and applied voltage verify that the ring studied in this paper is the inner ring.

The external ring is not observed when the laser excitation energy is well below the bandgap of the barrier material since such excitation typically does not generate the hole-rich area required for the external ring formation \cite{Butov02, Butov04, Rapaport04, Chen05, Haque06, Yang10}. On the contrary, the inner ring can be observed for the sub-barrier laser excitation, see Fig.~7 and Refs.~\cite{Ivanov06, Kuznetsova12}.

Figure~\ref{fig:8} shows the dependence of the inner ring on excitation power $P$. The ring radius slowly increases with $P$. The increase of $P$ results to the enhancement of the exciton density and, in turn, to a more effective screening of the in-plane disorder potential by excitons thus facilitating the transport of excitons over larger distances and increasing the inner ring radius. The increase of the inner ring radius with increasing excitation power is in agreement with the theory \cite{Ivanov06}.

Figure~\ref{fig:9} shows the dependence of the inner ring on applied voltage $V_{\rm g}$. The ring radius slowly increases with increasing $V_{\rm g}$. The applied voltage increases the lifetime of indirect excitons and, as a result, their density. Both a longer lifetime and a more effective screening of the disorder potential due to a higher exciton density facilitate the transport of excitons over larger distances thus increasing the inner ring radius.

\begin{figure}
\centering
\includegraphics[width=4.5cm]{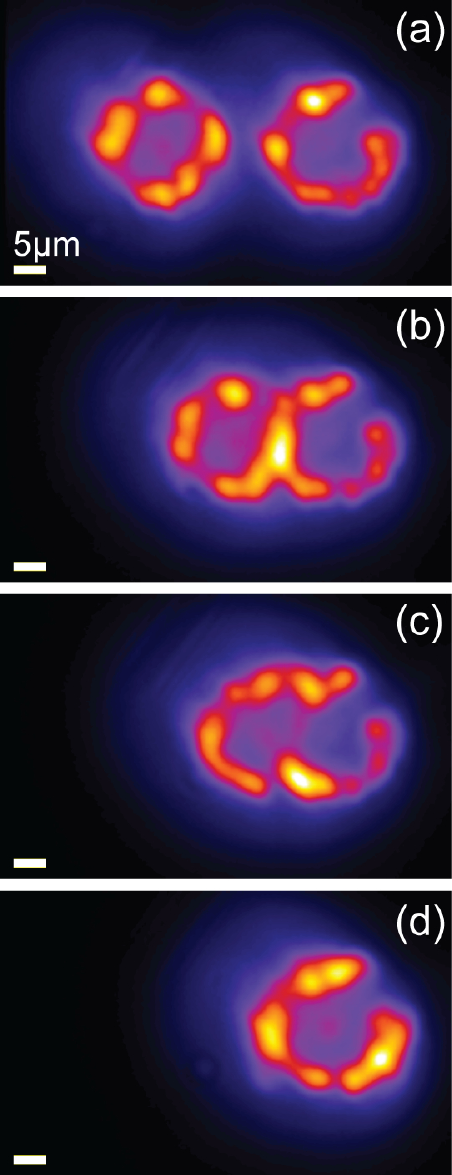}
\caption{Merging of two inner rings. (a-d) The emission pattern showing two inner rings created by two laser excitation spots merging into one spot as the left laser excitation spot is moved to the right. Laser power for both excitations $P=10\unit{\mu W}$, applied voltage $V_{\rm g}=-3\unit{V}$.}
\label{fig:12}
\end{figure}

For comparison, Figures~\ref{fig:10} and \ref{fig:11} show the dependence of the external ring on excitation power $P$ and applied voltage $V_{\rm g}$, respectively. The increase of $P$ increases the number of photogenerated holes thus increasing the external ring radius (Fig.~\ref{fig:10}), while the increase of $V_{\rm g}$ increases the number of electrically injected electrons thus reducing the external ring radius (Fig.~\ref{fig:11}) \cite{Butov04, Rapaport04, Chen05, Haque06, Yang10}. The external ring radius varies more quickly with $P$ and $V_{\rm g}$ than the inner ring radius, compare Fig.~\ref{fig:8} and \ref{fig:10}, \ref{fig:9} and \ref{fig:11}. Furthermore, there is a qualitative difference in the voltage dependence: The inner ring radius increases with $V_{\rm g}$ (Fig.~\ref{fig:9}) while the external ring radius reduces with $V_{\rm g}$ (Fig.~\ref{fig:11}).

A remarkable difference between the inner ring and external ring is also observed in the character of merging two rings. Figure \ref{fig:12} shows merging of two inner rings generated by two laser excitation spots as one excitation spot is moved towards another. Figure~\ref{fig:12}b shows that the intensity of the touching inner rings is added creating a bright region between the inner rings. This behavior is similar to the formation of the bright exciton cloud in the center of the optical trap in agreement with simulations \cite{Hammack06, Hammack07}. In contrast, merging of two external rings generated by two laser excitation spots is qualitatively different: As the spots are brought closer, the external rings attract one another, deform, and then open towards each other, forming a dark area between the external rings, see Fig.~2 in Ref.~\cite{Butov04}. This behavior, suggesting the existence of dark matter outside the external rings that mediates the interaction between the external rings, is in agreement with the transport model \cite{Butov04}. The model readily accounts for the attraction between the external rings, with this dark matter being just the electron flow outside each ring which is perturbed by the presence of another ring \cite{Butov04}.

\end{document}